\documentclass[letter]{aa} 

\usepackage{graphics}
\usepackage{longtable}
\usepackage{txfonts}
\usepackage{graphicx}
\usepackage{hyperref}

\begin{document}
\title{The binary Be star $\delta$ Scorpii at high spectral and spatial resolution}
\subtitle{II The circumstellar disk evolution after the periastron\thanks{Based on observations made with ESO telecopes at La Silla Paranal Observatory under GTO programme IDs 089.D-0105(A) and 089.D-0105(B)}~\thanks{Fig. ~\ref{visi2012} is only available in electronic form via
www.edpsciences.org}}

\authorrunning{A. Meilland et al.}

\titlerunning{$\delta$ Sco at high spectral and spatial resolution : The circumstellar disk evolution after the periastron}
   \author{A. Meilland \inst{1,2}, Ph. Stee \inst{1}, A. Spang\inst{1}, F. Malbet\inst{3}, F. Massi\inst{4}, and  D. Schertl\inst{5}}

   \offprints{ame@oca.eu}

\institute{Laboratoire Lagrange, UMR 7293 Universit\'e de Nice-Sophia Antipolis (UNS), Observatoire de la C\^ote d'Azur (OCA), Boulevard de l'Observatoire, B.P. 4229 F, 06304 Nice Cedex 4, France.
\and Physics and Astronomy Department, The University of Western Ontario, London, N6A 3K7, Ontario, Canada.
\and UJF-Grenoble 1 / CNRS-INSU, IPAG UMR 5274, Grenoble, F-38041, France
\and INAF – Osservatorio Astrofisico di Arcetri, Istituto Nazionale di Astrofisica, Largo E. Fermi 5, 50125 Firenze, Italy
\and Max-Planck-Institut f\"ur Radioastronomie, Auf dem H\"ugel 69, D-53121 Bonn, Germany}
  \date{Received; accepted }

\abstract{Classical Be stars are hot non-supergiant stars surrounded by a gaseous circumstellar disk that is responsible for the observed
infrared (IR) excess and emission lines. The influence of binarity on these phenomena remains controversial.}
{We followed the evolution of the environment surrounding the binary Be star $\delta$ Scorpii one year before and one year after the 2011 periastron to check for any evidence of a strong interaction between its companion and the primary circumstellar disk.}
{We used the VLTI/AMBER spectro-interferometric instrument operating in the K band in high (12000) spectral resolution to obtain information on both the disk geometry and kinematics. Observations were carried out in two emission lines: Br$\gamma$ (2.172\,$\mu$m) and  $\ion{He}{i}$ (2.056\,$\mu$m).}
{We detected some important changes in $\delta$ Scorpii's circumstellar disk geometry between the first observation made in April 2010 and the new observation made in June 2012. During the last two years the disk has grown at a mean velocity of 0.2\,km\,s$^{-1}$. This is compatible with the expansion velocity previously found during the 2001-2007 period. The disk was also found to be asymmetric at both epochs, but with a different morphology in 2010 and 2012.}
{Considering the available spectroscopic data showing that the main changes in the emission-line profiles occurred quickly during the periastron, it is probable that the differences between the 2010 and 2012 disk geometry seen in our interferometric data stem from a disk perturbation caused by the companion tidal effects. However, taking into account that no significant changes have occurred in the disk since the end of the 2011 observing season, it is difficult to understand how this induced inhomogeneity has been ``frozen'' in the disk for such a long period.}


   \maketitle
%

\section{Introduction} \par

We initiated an interferometric follow-up of the binary Be star $\delta$ Scorpii which has been regularly observed with the VLTI/AMBER (Petrov et al. 2007) and CHARA/VEGA (Mourard et al. 2009) instruments since 2007. In Meilland et al. (2011), hereafter Paper I, we managed to refine $\delta$ Scorpii's binary orbit and the orbital period was the only orbital element which was significantly different from the one determined by Tango et al. (2009). Consequently, we determined that the next periastron should take place around 2011 July 5 ($\pm$\,4\,days). This result was also confirmed by Tycner et al. (2011). Thanks to the combination of high spectral and spatial resolution we constrained the geometry and the kinematics of $\delta$ Scorpii's circumstellar environment. We confirmed that the line emission originates from a dense equatorial disk fully dominated by rotation. We found that the rotation appears to be Keplerian, with an inner boundary (photosphere/disk interface) rotating close to the critical velocity (V$_c$). The expansion was found to be approximately 0.2\,km\,s$^{-1}$.

\noindent With the measured  vsini of 175 km\,s$^{-1}$ (Berbacca \& Perinotto 1970), and the determined inclination angle of 28\,$\pm$\,8$^o$  we inferred that the star rotates between 60$\%$ and 105$\%$ of its critical velocity. Moreover, taking into account a possible underestimation of the vsini because of gravity darkening (Townsend et al. 2004), the lower limit on the stellar rotation would rise to about 70\,$\%$\,V$_c$. Finally, we measured the extension in the H$\alpha$ line, i.e., 4.8\,$\pm$\,1.5 mas, which is about the same as the binary separation at the periastron, i.e., 5.9\,mas. Thus, we suggested that a strong interaction between the already ejected matter and the companion should occur during the periastron passage.

\noindent This letter is organized as follows: in Sect. 2 we briefly summarize other recent observations of  $\delta$ Scorpii; in Sect. 3 we present our VLTI/AMBER observations and the data reduction. In Sect. 4, the data are analyzed and compared to those acquired in April 2010, one year before the periastron, and finally, a short conclusion is drawn in Sect. 5.

\section{Recent observations of $\delta$ Scorpii}

the star $\delta$ Scorpii (7 Scorpii, HR 5953, HD 143275) was observed by Millan-Gabet et al. (2010) using the CHARA/MIRC instrument in the H band. They resolved an elongated disk with a Gaussian FWHM of 1.18 x 0.91 mas. In the K band the disk was only marginally spatially resolved and they estimated an uncertain continuum disk FWHM of 0.7\,$\pm$\,0.3 mas. When taking into account the continuum/line flux ratio they obtained much larger sizes for the line emission regions, i.e., 2.2\,$\pm$\,0.4\,mas for  $\ion{He}{i}$ and 1.9\,$\pm$\,0.3\,mas for Br$\gamma$.  In Paper I,  we resolved the circumstellar disk in the H$\alpha$ (FWHM=\,4.8\,$\pm\,$\,1.5 mas), Br$\gamma$ (FWHM=2.9$\pm$0.5 mas), and the 2.06$\,\mu$m He$_{\rm I}$ (FWHM=\,2.4\,$\pm$\,0.3\,mas) lines, as well as in the K-band continuum (FWHM\,$\sim$\,2.4\,mas).

In a recent paper Che et al. (2012) imaged the disk distortion near the periastron. They revised the binary orbit and studied the primary disk properties during the periastron with NPOI and CHARA/MIRC interferometers. They were able to fit a global model where the secondary follows a Keplerian orbit, as already mentioned in Paper~I, and the disk properties were found to be stable through all their observations. The estimated the new periastron passage to be UT 2011 July 3 07:00\,$\pm$\,4:30 which is within the error bar of the prediction made in Paper~I. They also estimated the mass of the secondary, based on the revised binary orbit and radial velocity (RV) measurements from Miroshnichenko et al. (2001), to be $\sim$\,6\,M$_{\sun}$ with the primary mass 13.9\,M$_{\sun}$ estimated from v\,sin\,i, apparent Teff, and V-band photometry measurements. The mutual angles between the disk mid-plane and orbital plane were estimated to be either $\sim$\,119${\degr}$ or $\sim$\,171${\degr}$. From the fitted global model, they also found that $\sim$ 3\% of the H-band flux comes from a fully resolved envelope. The primary disk was found to be mainly symmetric and stable, contributing to 71.4\,$\pm$\,2.7\% of the total H-band flux. This implies a quiescent inner disk and no ongoing material outflow after the periastron which is very different from what we obtained in the K band.

On the other hand, Rivinius et al. (2012) conducted a spectroscopic study of $\delta$ Scorpii emission lines during the periastron. They found that H$\alpha$ and H$\beta$ profiles were strongly modified at the periastron. The major modification was that the Violet/Red (V/R) ratio went from V/R$>$1 to V/R$<$1 in only a few days. They argued that this was due to the direct tidal effect of the companion on the primary disk. 

We note that we looked at many H$\alpha$ and H$\beta$ observations made by amateurs available in the BeSS database \footnote{ http://basebe.obspm.fr} and that the line profiles did not change significantly from the state detected by Rivinius et al. (2012) after the periastron.

\begin{figure}[!b]
       \centering  
       \includegraphics[width=0.45\textwidth]{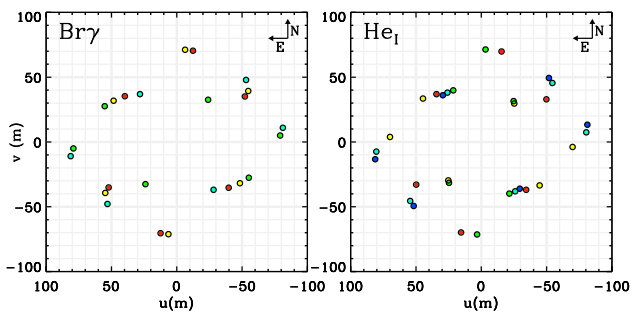}
     
\caption{$\delta$ Scorpii (u,v) 49lan coverage obtained  in 2012  with VLTI/AMBER observations centered on Br$\gamma$ (left) and 2.06\,$\mu$m $\ion{He}{i}$ (right) emission lines.}
\label{uvplan}
\end{figure}

\section{Observations and data reduction}

The star $\delta$ Scorpii was observed in June 2012 during two half-nights with the VLTI 1.6m auxiliary telescopes (AT) using the AMBER beam recombiner (Petrov et al. 2007). The observations were carried out using high spectral resolution mode (R=12000) centered on either Br$\gamma$ or $\ion{He}{i}$ (2.06\,$\mu$m) for each of the two half-nights. In order to enhance the signal-over-noise ratio (S/N) by enabling longer exposure time, the fringe tracker FINITO was also used. The star HD\,139663 (D\,=\,2.133\,$\pm$\,0.147\,mas, Lafrasse et al. 2010) was used as an interferometric calibrator.

The data were reduced using the standard AMBER data reduction software \texttt{amdlib v3.0.3b1} (see Tatulli et al. 2007 and Chelli et al. 2009 for detailed information on the AMBER data reduction). The average-raw complex visibility and closure phase was determined using the standard method, i.e., keeping the 20\,\% of the frames with the higher S/N ratio. The interferometric calibration was then done using custom-made scripts\footnote{available at \url{https://www.oca.eu/spip.php?article562}} described in Millour et  al. (2008). The (u,v) plan coverage for all the observations is plotted in Fig.~\ref{uvplan}.

\begin{table}
\caption{ VLTI/AMBER log for the 2012 $\delta$ Scorpii observations.}
{\centering \begin{tabular}{ccccc}
\hline  Observation& Triplet & \multicolumn{2}{c}{Projected Baselines}\\
  ~~~~~~Date ~~~~~ Time&  & L. (m) & P.A.($^o$)  \\
\hline
\hline
\multicolumn{4}{c}{Observations centered at 2.17\,$\mu$m (Br$\gamma$)}\\\hline
2012-06-22 23:46&D0-H0-G1& 53/ 71/ 63&  48/ 170/ 124\\
2012-06-23 00:32&D0-H0-G1& 57/ 71/ 67&  57/ 175/ 126\\
2012-06-23 01:21&D0-I1-H0& 79/ 40/ 62&  94/ -36/  63\\
2012-06-23 02:05&D0-I1-G1& 82/ 46/ 71&  98/-143/ 132\\\hline
\multicolumn{4}{c}{Observations centered at 2.06\,$\mu$m (He$_I$)}\\\hline
2012-06-23 23:16&D0-H0-G1& 50/ 71/ 60&  43/ 167/ 123\\
2012-06-24 00:07&D0-I1-H0& 70/ 39/ 56&  87/ -40/  53\\
2012-06-24 00:52&H0-I1-G1& 40/ 45/ 71& 142/-152/ 177\\
2012-06-24 01:35&D0-I1-G1& 80/ 46/ 71&  95/-146/ 130\\
2012-06-24 02:18&D0-I1-G1& 82/ 47/ 71&  99/-141/ 134\\
\hline
\end{tabular}\par}

\end{table}

\section{The disk geometry and kinematics}

\subsection{Qualitative comparison of the 2010 and 2012 data}

\begin{figure*}[!t]
       \centering  
       \includegraphics[width=0.955\textwidth]{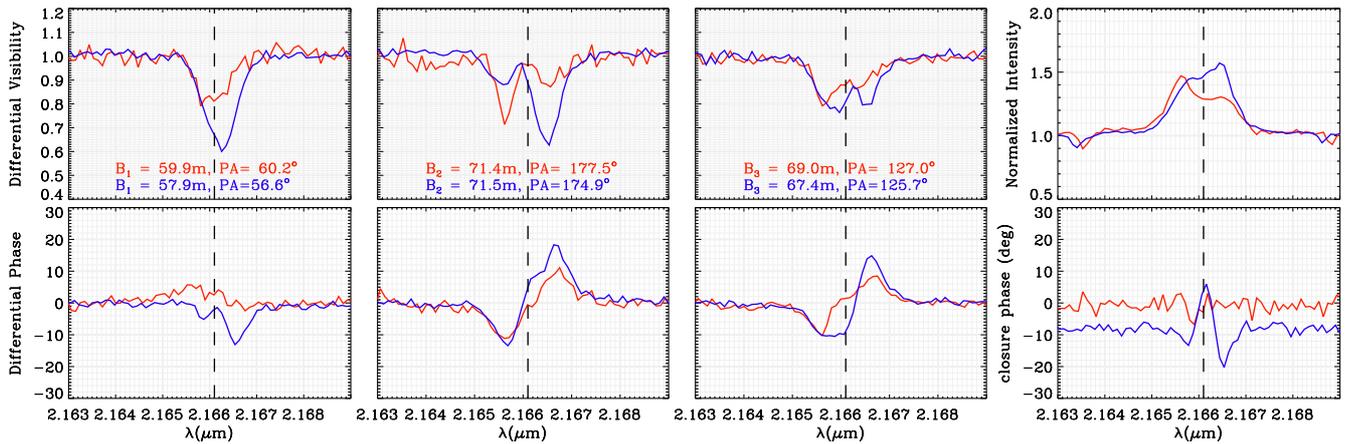}
     
\caption{2010 (in red) and 2012 (in blue)  Br$\gamma$  VLTI/AMBER data for a similar projected triplet of baselines.}
\label{visi2010_2012}
\end{figure*}

The Br$\gamma$ and $\ion{He}{i}$ absolute visibility  show a strong variability with an amplitude of about 0.2. Assuming that the binary separation at the epoch of the observations should be 90.8\,mas with a position angle (PA) of 156.08$^o$, according to Paper I orbital parameters, these variations are compatible both in terms of amplitude and frequency with the binary modulation. However, the accuracy of the absolute visibility of VLTI/AMBER data taken in HR mode is not sufficient to significantly refine the orbital parameters.

\begin{figure*}[!t]
       \centering  
       Br$\gamma$ differential visibilities differential phases, and closure phases   
       \vspace{0.3cm} 
       \includegraphics[width=0.8\textwidth]{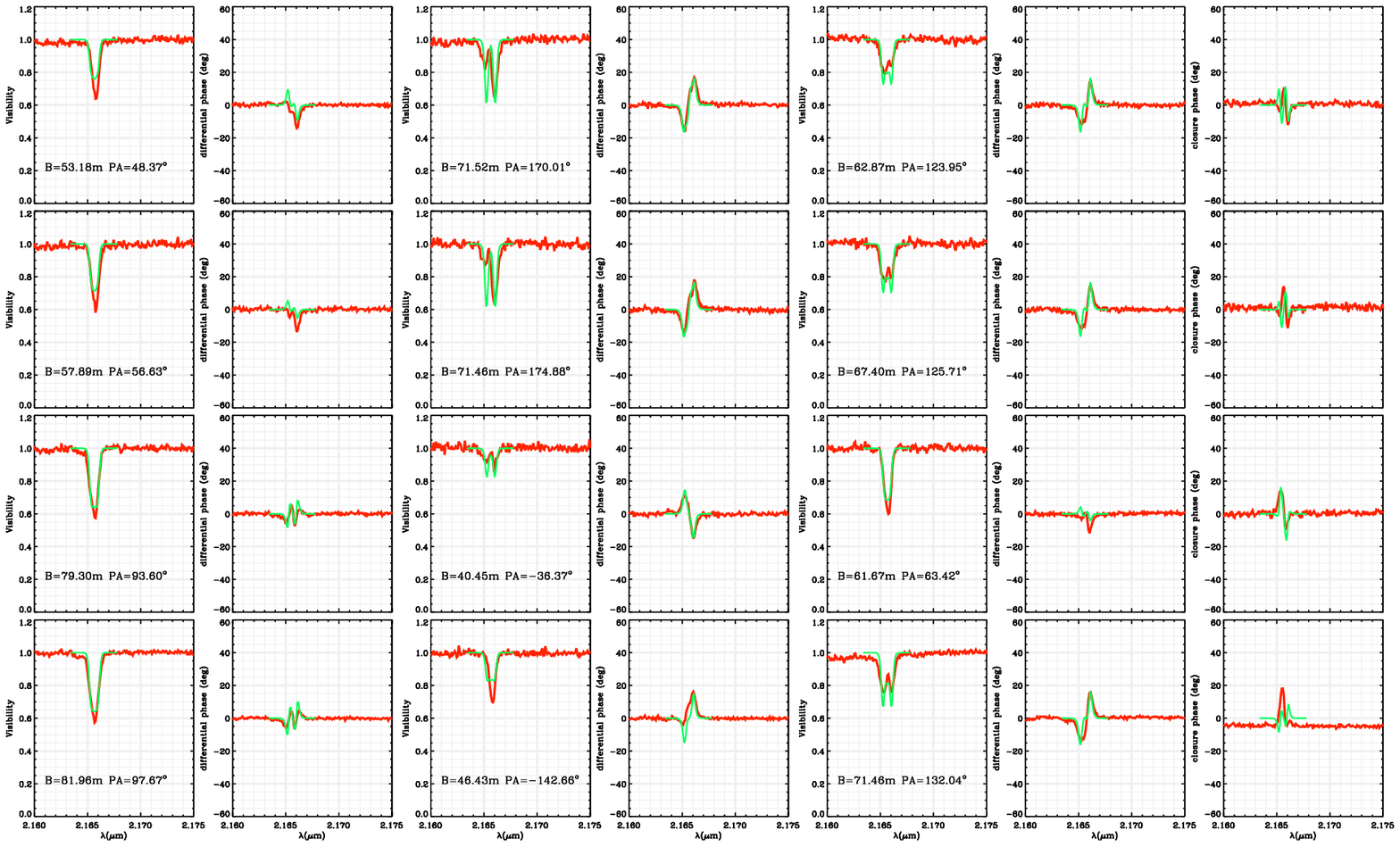}
       \vspace{0.3cm}\\
			He$_I$ differential visibilities differential phases, and closure phases
			\vspace{0.3cm}
       \includegraphics[width=0.8\textwidth]{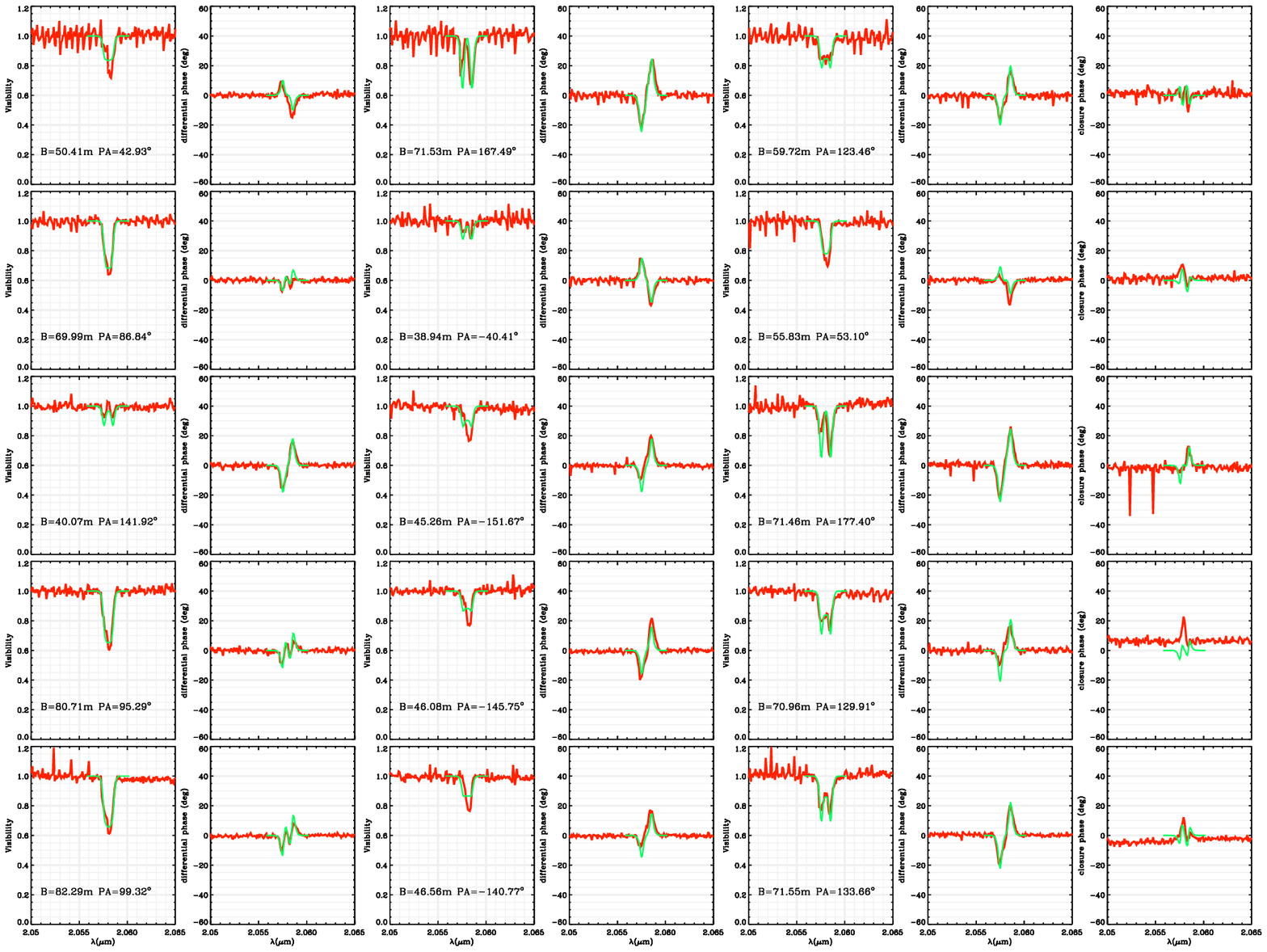}
       \vspace{0.5cm}
       
       \centering  

         \includegraphics[width=0.40\textwidth]{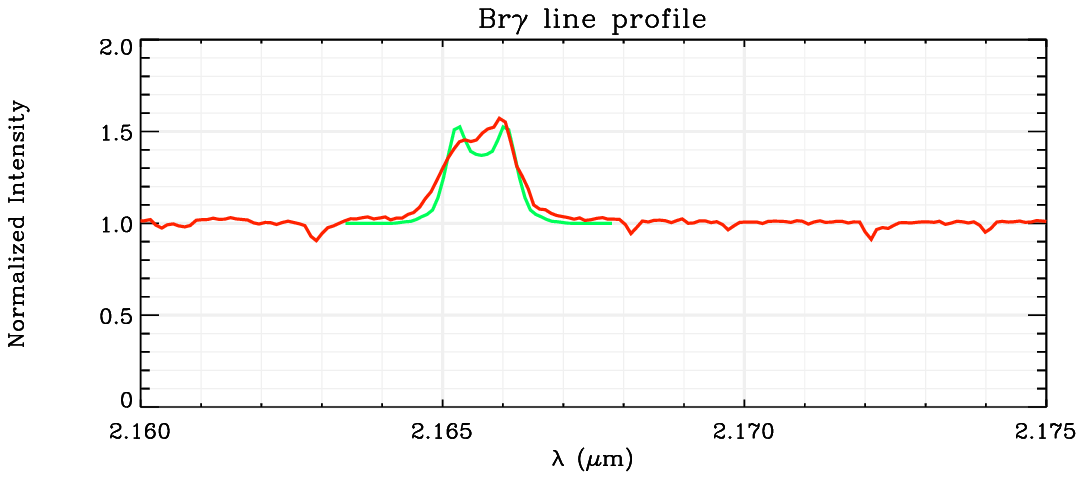}
          \includegraphics[width=0.40\textwidth]{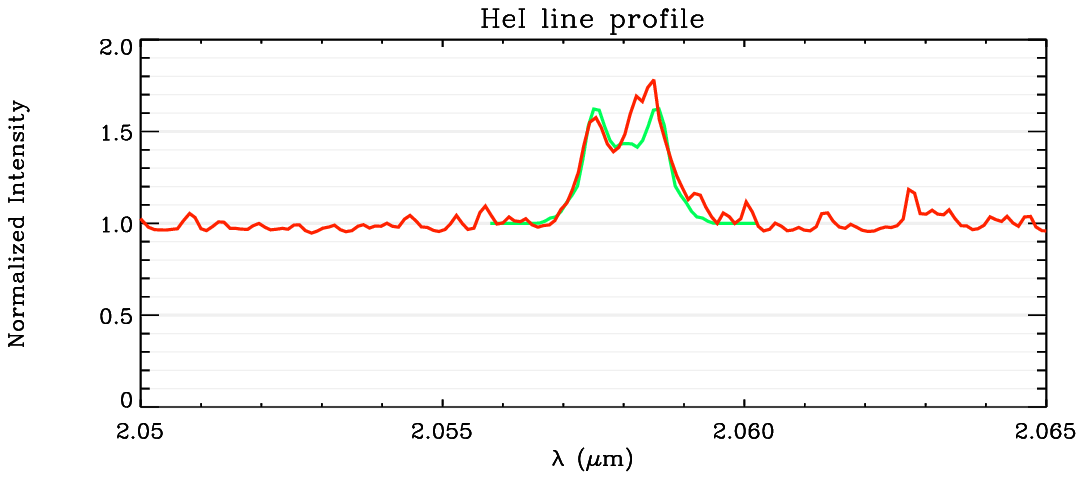}
\caption{The full dataset for $\delta$ Sorpii 2012 VLTI/AMBER observations. The red lines represents the VLTI/AMBER data, and the green ones, the best-fit symetric kinematic model.}
\label{visi2012}
\end{figure*}

The differential visibilities and phases, closure phase, and line profile from the 2010 Br$\gamma$ observation are plotted  in Fig.~\ref{visi2010_2012}. In the same figure, one single Br$\gamma$ 2012 observation with similar projected baselines is overplotted. The full 2012 dataset for Br$\gamma$ and $\ion{He}{i}$ emission lines is presented in Fig.~\ref{visi2012} (online).

The visibility and phase variations shown in Fig.~\ref{visi2010_2012} are typical for a rotating disk. As discussed in detail in Meilland et al. (2012), the morphology of visibility variations for a purely rotating disk depends on the baseline orientation. Considering a non-fully resolving baseline, the variation is V-shaped for a baseline aligned with the disk's minor axis (baseline 1 of Fig~\ref{visi2010_2012}) and W-shaped for a baseline aligned with the major axis (baseline 2 of Fig~\ref{visi2010_2012}). Baselines with intermediate orientation exhibit intermediate morphologies (see, for example, the weak W-shaped variation for the third baseline shown in Fig.~\ref{visi2010_2012}).

The S-shaped phase variations also strongly depend on the baseline orientation. In the case of a purely rotating disk, the amplitude is maximum for a baseline aligned with the major axis and zero for a perpendicular baseline. Thus, from the simple analysis of data shown in Fig.~\ref{visi2010_2012} it seems that the disk orientation did not significantly change between 2010 and 2012 with the major axis being oriented roughly in the direction North-South.

However, Fig.~\ref{visi2010_2012} also clearly shows important changes in the visibilities, phases, and line profile between 2010 and 2012. Some of them favor the hypothesis of a significant growth of the equatorial disk between the two epochs:

\begin{itemize}
\item The visibility drop is clearly deeper in 2012.
\item The 2012 S-shaped phase variations are larger.
\item The 2012 line profile equivalent width (EW) is larger. 
\item the double-peak separation is narrower in 2012.
\end{itemize}

Nevertheless, other major changes seen in Fig.~\ref{visi2010_2012} are not related to a growth of the disk:

\begin{itemize}
\item The V/R ratio went from 1.2 in 2010 to 0.9 in 2012.
\item The W-shaped-visibilities peak ratio is reversed.
\item The 2012 S-shaped phase variations are asymmetric.
\end{itemize}

Such changes of asymmetries in the spectrum, visibilities, and phases across the Br$\gamma$ line, are clear evidence of changes in the circumstellar matter distribution around the star. In the case of a rotating disk, these asymmetries usually stems from the inhomogeneity of the disk. The origin and nature of the disk's inhomogeneities and their evolution around the periastron are discussed in Sect. 4.3.

We note that all this evidence of the evolution of the disk around $\delta$~Scorpii between 2010 and 2012 are seen not only in the data plotted in Fig.~\ref{visi2010_2012}, but also on all the Br$\gamma$ and  $\ion{He}{i}$ dataset (see Fig.~\ref{visi2012} online).

\subsection{Modeling of the 2012 data using the kinematic model}

To model the wavelength dependence of the visibility, differential phase, and closure phase in the observed emission lines we used the simple kinematic model developed for fast fitting of an expanding and/or rotating thin equatorial disk already described in paper I. The model is presented in more detail in Delaa et al. (2011) and Meilland et al. (2012).

We used the best-fit model parameters of the 2010 data from Paper I as a starting model. As expected, we obtained a very unsatisfactory fit with a $\chi^2_r$ of 24.7. We then assumed that the stellar and kinematics parameters should not change between 2010 and 2012, and that only the Br$\gamma$ and $\ion{He}{i}$ disk extension and line equivalent width (EW) should have evolved. In this case we found a best-fit model for FWHM$_{Br\gamma}$\,=\,8.0\,$\pm$\,0.5\,R$_\star$ (4.2\,$\pm$\,0.3\,mas), EW$_{Br\gamma}$\,=\,7.2\,$\pm$\,0.3\,$\AA$,   FWHM$_{HeI}$\,=\,5.2\,$\pm$\,0.3\,R$_\star$ (2.8\,$\pm$\,0.2\,mas), and EW$_{HeI}$\,=\,9.1\,$\pm$\,0.3\,$\AA$. However, for this set of values the $\chi^2_r$ is still high, i.e., 17.4.  

We then performed a fit after releasing constraints on the stellar inclination angle (i), disk major axis orientation (PA), and disk kinematics, i.e.,\textsl{} rotational velocity at the photosphere (V$_{rot}$) and exponent of the rotational law in the disk ($\beta$). Thus, the only fixed parameters for this set of models are the distance taken from van Leeuwen (2007) and the stellar radius taken from Paper I. The best-fit model parameters are presented in Table~\ref{model_params} and the model visibility, phases, and line profiles are overplotted on Fig.~\ref{visi2012} (online only).

The $\chi^2_r$ for this best-fit model, however, is still high, i.e., 14.6. By separating the fit on each line we found that the $\ion{He}{i}$ data  have a significantly better fit (i.e., $\chi^2_r$\,=\,8.6) than the Br$\gamma$ data ($\chi^2_r$\,=\,22.7). For both lines, the largest deviations between the observations and the data comes from the asymmetrical shape of the line profile as well as some visibility and phase variations. This effect will be discussed in the next section.

\begin{table}[!b]
\caption{The 2010 and 2012 best-fit kinematic models.\label{model_params}}
\centering \begin{tabular}{cccc}
\hline
~~~~Parameter	~~~~& ~~~~(unit)~~~~				& ~~~~~~~~2010~~~~~~~~				&~~~~~~~~2012~~~~~~~~	\\
\hline\hline
\multicolumn{4}{c}{\textbf{Global geometric parameters}}\\
$R_\star$ (fixed)			&	(R$_\odot$)				&	8.5											& 8.5\\
d			(fixed)			&(pc)								&	150											& 150 \\
i	   								 	&(deg)							& 30\,$\pm$\,8 							& 26\,$\pm$\,5 \\
PA 										&(deg)							& -12\,$\pm$\,7							& -14\,$\pm$\,4\\
\hline
\multicolumn{4}{c}{\textbf{Global kinematic parameters}}\\
V$_\mathrm{rot}$ 	&(km.s$^{-1}$	)			&	500\,$\pm$\,50							& 500\,$\pm$\,50\\
$\beta$								&										&	0.5\,$\pm$\,0.1							& 0.5\,$\pm$\,0.05\\
\hline
\multicolumn{4}{c}{\textbf{Br$\gamma$ disk geometry}}\\
FWHM$_\mathrm{Br\gamma}$ &(R$_\star$)				&	5.5\,$\pm$\,1								&7.8\,$\pm$\,0.3\\
EW$_\mathrm{Br\gamma}$&($\AA$)		  			&	6.5\,$\pm$\,0.5							&6.2\,$\pm$\,0.2\\
\hline
\multicolumn{4}{c}{\textbf{He$_I$ disk geometry}}\\
FWHM$_\mathrm{HeI}$			&(R$_\star$)				&	4.5\,$\pm$\,0.5							&4.9\,$\pm$\,0.2\\
EW$_\mathrm{HeI}$			&($\AA$)		  			&	8.5\,$\pm$\,0.5							&8.8\,$\pm$\,0.3\\
\hline
$\chi^2_r$						&		  							&	6.2																		&14.6\\
\hline\hline
\end{tabular}
\end{table}

We also note a second source of deviation. Our best-fit model clearly underestimates the visibility for baselines close to the disk's minor axis. However, using a larger disk extension, we would overestimate the visibility for the other baselines and the over-resolution of the disk would  strongly modify the shape of the phase variations. Such underestimation of the minor-axis visibility was already detected for $\alpha$ Col, one of the Be stars observed with VLTI/AMBER (Meilland et al. 2012). However, its origins remain unknown.

\subsection{The disk asymmetry}

Okazaki (1997) proposed that the cyclic variation of the V/R peak ratio with a period of several years originates in the precession of one-armed density waves in the disk.  This assumption was compatible with spectro-interferometric observations of $\gamma$~Cas (Berio et al. 1999), $\kappa$~CMa (Meilland et al. 2007), and $\zeta$~Tau (Caricofi et al. 2009). This could also explain the change of the peaks ratio of the W-shaped visibility seen in Fig.~\ref{visi2010_2012}. However, according to observations made by Rivinus et al. (2012), the V/R changes for $\delta$ Scorpii occurred during a very short timescale of a few weeks around the periastron. Thus, it seems that the hypothesis of the one-arm oscillation cannot explain the current disk properties.

We also note that, using our simple axisymmetric model, we obtained a better fit on the $\ion{He}{i}$ data than on the Br$\gamma$ data. Considering that the Br$\gamma$ emission line is formed further from the primary, it is supposed to be more perturbed by the companion tidal forces than the $\ion{He}{i}$ line. Consequently, the worse fit of the Br$\gamma$ data might be a clue of a more important tidal perturbation. One can also argue that the broad-band measurements taken by Che et al. (2012) could not detect any tidal effect as the H-band continuum stems from an even smaller part of the disk than the $\ion{He}{i}$ line. Nevertheless, in the frame of an asymmetry induced by tidal effects in June-July 2011, it is also difficult to understand why the V/R remained almost constant between July 2011 and our observations in June 2012. One should expect the induced inhomogeneity to be either in co-rotation or precession, causing cyclical V/R variations with timescales ranging from weeks to years that are not seen in the available spectroscopic data.

\subsection{Evolution of the disk extension}

Our VLTI/AMBER data clearly gives evidence that the Br$\gamma$ emission FWHM increased by about 40$\%$ (i.e., 2.3\,$\pm$\,1.3\,R$_\star$) between 2010 and 2012. Assuming that this was directly cause by a physical growth of the circumstellar disk we derived a mean expansion velocity of 0.20$\pm$0.11\,km\,s$^{-1}$. This value is similar to  the one previously determined for $\delta$ Scorpii by Miroshnichenko et al. (2003) for the 2001-2003 period, i.e., 0.4\,km\,s$^{-1}$, and the one derived in Paper I for the 2000-2005 and 2005-2007 periods, 0.24\,km\,s$^{-1}$ and 0.19\,km\,s$^{-1}$, respectively. It is also compatible with the one determined by Kanaan et al. (2008) for the transient disk surrounding the binary Be star Achernar, i.e., 0.27\,$\pm$\,0.08\,km\,s$^{-1}$.

The $\ion{He}{i}$ line, however, did not exhibit the same variation as the Br$\gamma$ line. Its growth, of about 8$\%$, remains within the uncertainties of the derived extension. One possible explanation for this difference is that $\ion{He}{i}$ has a higher excitation energy. Consequently, it is produced by denser and hotter material than Br$\gamma$. Thus, even if the disk is increasing in size as a function of time the $\ion{He}{i}$ emitting region should remain compact, whereas the Br$\gamma$ may be produced in the newly formed outer part of the disk.

In a recent paper, Stee et al. (2012) measured the disk extension in these two lines for the relatively steady disk of the Be star HD110432. The values derived for their measurement using the same methods showed that the $\ion{He}{i}$ emission is about 25$\%$ smaller than the Br$\gamma$ emission. This value is between the one we found for $\delta$ Scorpii for our 2010 (18$\%$) and 2012 (37$\%$) models.

\section{Conclusion}

Using the unique high spectral and spatial resolutions of the VLTI/AMBER instrument, we studied the evolution of $\delta$~Scorpii's circumstellar environment one year before and one year after the 2011 periastron passage. At both epochs, the data clearly shows that the ejected matter is located in a thin equatorial disk in Keplerian rotation as it is for other observed Be stars. Since 2000, the disk seems to have grown with a quasi-constant velocity of about 0.2\,km\,s$^{-1}$ and was still growing during the 2010-2012 period. Such speed is compatible with the viscous excretion disk model proposed by Lee et al. (1991).

We found evidence that the disk was asymmetric at both epochs of our observations, but with a different morphology in 2010 and 2012. According to spectroscopic observations made by Rivinius et al. (2012), the major modifications occurred near the periastron. Thus, the observed asymmetry cannot originate from a one-arm oscillation as proposed by Okazaki (1997), but must stem from tidal effects induced by $\delta$ Scorpii's companion. However, according to available spectroscopic data and our 2012 observations, the disk asymmetry seems to be ``frozen'' since the periastron. Finally, long-term spectroscopic and interferometric follow-up are needed to understand both the nature of the inhomogeneity after the periastron and its link with the tidal effects of the companion.

\begin{acknowledgements}
The Programme National de Physique Stellaire (PNPS) and the Institut National en Sciences de l'Univers (INSU) are acknowledged for their financial supports.   A. Meilland acknowledge financial support from the University of Western Ontario. The authors would like to thank ESO staff for their help to make the observations a success and more particularly J-B Lebouquin and F. Rantakyro. 
\end{acknowledgements}

\end{document}